\documentclass[pre,twocolumn,groupedaddress,showpacs,showkeys]{revtex4}

	\usepackage{amssymb}
	\usepackage{amsmath}
	\usepackage{graphicx}
	\usepackage{subfigure}
	\usepackage{dsfont}
	\usepackage{mathrsfs}
	\usepackage{natbib}       
	\usepackage[dvips]{color} 
\newcommand{\fix}[1]			
             {{\color{magenta} #1 }}

\newcommand{\beq}{\begin{equation}}
\newcommand{\eeq}{\end{equation}}
\newcommand{\ee}[1] {\label{#1} \end{equation}}
\newcommand{\bea}{\begin{eqnarray}}

\newcommand{\eea}{\end{eqnarray}}

\begin{document}
\newcommand{\changeA}[1]{{\large\bf #1}}  
\newcommand{\changeO}[1]{{\large\bf #1}}  
                                
\title{
Fractal dimension of domain walls in two-dimensional Ising spin glasses
}
\author{
O.~Melchert$^1$ and A.~K.~Hartmann$^2$
}
\affiliation{
$^1$Institut f\"ur Theoretische Physik, Universit\"at G\"ottingen, 37077 G\"ottingen, Germany\\
$^2$Institut f\"ur Physik, Universit\"at Oldenburg, 26111 Oldenburg, Germany
}
\date{\today} 

\begin{abstract}
We study domain walls in 2d Ising spin glasses
in terms of a minimum-weight path problem. Using this approach, large
systems can be treated exactly. Our focus is on the 
 fractal dimension $d_f$ of domain walls, which describes via 
 $\langle \ell \rangle\!\sim\!L^{d_f}$ the growth of the average
 domain-wall length with 
systems size $L\times L$. 
Exploring systems up to $L\!=\!320$ we yield $d_f\!=\!1.274(2)$ for
the case of Gaussian disorder, i.e.\ a much higher accuracy
compared to previous studies. 
For the case of bimodal disorder, where many equivalent domain walls
exist due to the degeneracy of this model, we obtain a true lower bound
$d_f\!=\!1.095(2)$ and a (lower) estimate $d_f\!=\!1.395(3)$ as upper bound.
Furthermore, we study the distributions of the domain-wall
lengths. Their scaling with system size can be described also only by the
exponent $d_f$, i.e.\ the distributions are monofractal. 
Finally, we investigate the growth of the domain-wall
width with system size (``roughness'') and find a linear behavior. 
\end{abstract} 


\maketitle

\section{Introduction}

In this paper we examine the scaling behavior of minimum-energy 
domain wall (MEDW)  
excitations for  two-dimensional Ising spin glasses. Spin glasses are
a prototypical model
\cite{binder1986,mezard1987,fischer1991,young1998} 
for systems with quenched disorder in statistical mechanics. 
In general, despite more than two
decades of intensive research, many properties of spin glasses,
especially in finite dimensions, are still not well understood. 
The situation is slightly better for two-dimensional spin glasses,
because  it is now widely accepted that no
ordered phase for finite temperatures exists 
\cite{rieger1996,kawashima1997,hartmann2001,houdayer2001,carter2002}.
For $d=2$ the behavior can be described well by a  zero-temperature  
droplet scaling (DS) approach \cite{mcmillan1984,fisher1988,fisher1986}.
Within this approach, the excitation energy of domain walls and other
excitations 
scales like $\Delta E \sim L^{\theta}$, where $L$ is the system size
and $\theta$ is referred to as stiffness exponent. Within DS, $\theta$
is assumed to be universal for all types of excitations. 
Further, the
surface of an excitation is assumed to display a scaling that is
characterized by a fractal dimension $d_f$. 

For Gaussian disorder of the interactions, prior investigations of
domain walls  resulted in
estimates for the stiffness exponent $\theta = -0.287(4)$
\cite{hartmann2001,aspect-ratio2002} and the values for the MEDW fractal dimension
listed in Tab.\ 
\ref{t:fract_gauss}. For this model, it was recently indeed confirmed
\cite{droplets2003,droplets_long2004}
that the value of $\theta$ is the same also for other types of excitations.  
Recent studies suggested that such domain walls are governed by
stochastic Loewner evolution (SLE) processes
\cite{amoruso2006,bernard2006}. 
Within conformal-field theory, it seems
possible to relate the MEDW fractal dimension $d_f$ to the 
stiffness exponent $\theta$ via
\begin{equation}
  \label{eq:exponents}
d_f-1 = 3/[4(3+\theta)]\,.
\end{equation}
Note that for a wide range of values of $\theta$, one gets 
very similar results for the fractal dimension, e.g. 
$d_f(\theta=-0.2)=1.268$,
while $d_f(\theta=-0.3)=1.278$. Hence, a high accuracy is needed
to verify whether the proposed relation is compatible with the data.
Note that the error bars of the previous results for $d_f$ are 
typically 
of order $10^{-2}$ or larger, i.e. ten times larger than our 
high-precision result. Also, we reach much larger system sizes
than previous work (also a bit larger than the recent 
study in Ref.\ \cite{weigel2007}, which is anyway for 
a different lattice type), which also  reduces  systematic
	errors due to  unknown corrections to scaling.

\begin{table}[thb]
\begin{center}
\begin{tabular}[c]{l@{\quad}l@{\quad}l@{\quad}l@{\quad}l}
\hline
\hline
Reference                                       & $d_f$         & Geom. &  System               &  Alg. \\
\hline
Middleton \cite{middleton2001}                  & $1.25(1)$     & tr & $256\times 256$  & M \\
Bray/Moore \cite{bray1987}                     & $1.26(3)$     & sq & $12\times 13$    & TM \\
Kawashima/Aoki \cite{kawashima1999}            & $1.28(2)$     & sq & $48\times 48$            & M \\
Bernard \emph{et.\ al.\ }\cite{bernard2006}     & $1.28(1)$     & tr & $720\times 360$  & M \\
Rieger \emph{et.\ al.\ } \cite{rieger1996}      & $1.34(10)$    & sq & $30\times 30$            & BC \\
Palassini/Young      \cite{palassini1999}       & $1.30(8)$     & sq & $30\times 30$ &  BC \\
Weigel/Johnston \cite{weigel2007}		& $1.273(3)$	& hex & $256\times 256$ & M \\ 
\hline
\hline
\end{tabular}
\end{center}
\caption{{
Previous results on the fractal dimension of MEDWs arising from Gaussian
disorder in two dimensions. From left to right: Reference, estimate
value of the fractal dimension $d_f$, 
geometry of the system (sq: square lattice, tr: triangular lattice, hex: hexagonal lattice),
largest system studied and numerical algorithm used, for details see
cited references (M = matching
approaches, TM = ($T=0$) transfer matrix method, 
BC = branch-and-cut algorithm).
}\label{t:fract_gauss}}
\end{table}

\begin{figure}[t]
\centerline{
\includegraphics[width=0.95\linewidth]{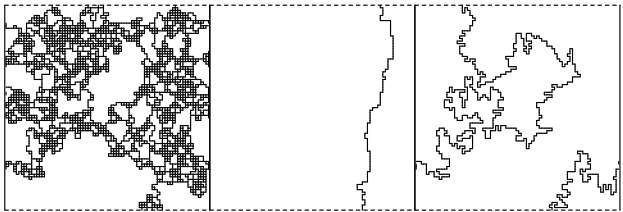}} 
\caption{
For the $\pm J$ model, a high degeneracy appears. Hence, for each of the GSs, 
many MEDWs separating the left and right border exist, which having all the 
same minimum energy. 
Left: many MEDWS for a sample system of size $L=80$, middle (right): corresponding $\pm J$--min ($\pm J$--max) MEDW. 
Dashed lines denote free boundary conditions, solid lines indicate periodic boundary conditions. 
 \label{fig:sampleDWs}}
\end{figure}

For the $\pm J$--model it was found that the MEDW energy saturates at a
nonzero value for large system sizes \cite{hartmann2001,amoruso2003},
which means $\theta=0$. 
It exhibits a high degeneracy of ground states (GSs) and thus allows
for numerous MEDWs, e.g.\ varying in length, see
Fig.\ \ref{fig:sampleDWs}. As a result, the
concept of a MEDW is not clear-cut. 
Referring to this model, the SLE scaling relation above cannot be adopted.  
Recent attempts to capture the fractal properties of typical MEDWs
arising from bimodal disorder concluded with values $d_f=1.30(1)$
\cite{roma2006} and $d_f=1.283(11)$ \cite{weigel2007}. Nevertheless,
in these works there is still
no systematics concerning the selection of a representative MEDW, 
as mentioned in Ref. \cite{weigel2007}.
The
sampling of the domain walls is not controlled, hence
different configurations having the same energy do {\em not}
contribute to the results with the same weight. Currently, no fast algorithm
is known which allows to sample the degenerate GSs and/or MEDWs
correctly. For a correct sampling, all degenerate configurations must
contribute to the results with
the same weight or probability. So far this can be done only 
for small systems through enumeration of all
GS configurations \cite{landry2002}.
  Some investigations use a scaling relation from droplet theory
\cite{fisher1988}, which quotes that the variance of the DW entropy
should also scale like a power law with the system size, with the
same exponent $d_f$. In a recent investigation based on this relation
for
square lattices of size up to $ L=256$ \cite{aromsawa2007} a value of 
$d_f=1.090(8)$ for $24\le L\le96$ and $d_f=1.30(3)$ for $L\ge 96$ was reported.
On the other hand,
other studies on the scaling behavior of the MEDW entropy
\cite{fisch2006,fisch2007} suggested that the behavior of zero-energy 
DWs is not
consistent with the droplet scaling picture. In this context it was
proposed to treat MEDWs of zero and non zero energies as distinct
classes. These findings were supported by calculations for square
lattices as well as for an aspect ratio different from unity.
Hence, due to these results,
a direct determination of the MEDW fractal dimension appears to be preferable.

So as to shed further light on their fractal properties, we
investigate MEDWs originating from bimodal and Gaussian disorder.
 We use  exact algorithms \cite{opt-phys2001}, which allow to
obtain the domain walls with the true lowest energy.
Here, we describe a novel approach to the problem of finding MEDWs in terms 
of minimum-weight paths, detailed in the forthcoming section. 
It allows to put a lower bound and an upper estimate on the scaling behavior of MEDWs 
and it might further provide a sound basis to investigate the scaling behavior 
of typical MEDWs for the $\pm J$ model, similar to the
random-field Ising model, where all ground states can be represented
by a certain graph \cite{daff21998}, which should allow for unbiased sampling
of ground states.
Moreover, the corresponding picture of minimum-weight paths on 
undirected graphs,  
that allow for negative edge weights, leads to a percolation problem 
that appears to be  interesting 
on its own, i.e.\ seems to be in a new universality class \cite{neg_percol}. 
For the bimodal system, to
avoid aforementioned problems of sampling correctly 
equivalent MEDWs of different length, we
ask for those MEDWs with extremal lengths, bearing a true lower bound and an
upper estimate $1.095(2)<d_f<1.395(3)$ 
in case of bimodal disorder. In addition, for the Gaussian 
system, we study square lattices with a large number of samples and 
much larger system sizes than
previously studied in the literature, yielding a  result
for the fractal dimension $d_f=1.274(2)$, 
which has an enhanced precision compared to the results shown in 
Tab.\ \ref{t:fract_gauss} and 
compares well with the value $d_f^{\mathrm{SLE}}=1.2764(4)$ obtained 
via inserting
$\theta=-0.287(4)$ into Eq.\ (\ref{eq:exponents}).
Finally, we also study for the first time the scaling-behavior
of the domain-wall roughness and of the distributions of domain-wall
length and width.

The paper is organized as follows. In the next section, we present the
details of the model and the numerical algorithms we have applied. In
the third section, we show all our results. We conclude with a summary
in section four. 


\section{Model and Method}
\label{sec:model}

In the framework of this paper, we present GS
calculations of two-dimensional Ising spin glasses with nearest-neighbor
interactions. The model consists of $N=L \times L$ spins
$\sigma=(\sigma_1,\ldots,\sigma_N)$ with $\sigma_i=\pm
1$ located on the sites of a square lattice. The energy is given by
 the Edwards-Anderson Hamiltonian 
\begin{eqnarray}
H(\sigma) = -\sum_{\langle i,j \rangle} J_{ij} ~\sigma_i \sigma_j,
\end{eqnarray}
where the sum runs over all pairs of neighboring spins, with
periodic boundary conditions (BCs) in the $x-$direction and open BCs
in the $y-$direction. The bonds
$J_{ij}$ are quenched random variables drawn
according to a given disorder distribution. They can take either sign
and thus lead to competing interactions among the spins. Here, we consider
two kinds of disorder
distributions: (i) Gaussian with zero mean and variance one,
and
(ii) a bimodal distribution $J_{ij}=\pm 1$ with equal probability
($\pm J$ model).
 
In this work, we study minimum-energy domain walls, 
which are certain excitations that are defined,
for each realization of the disorder, relative to two spin
configurations: $\sigma$ is a ground state with respect
to periodic BCs ($x-$direction) and $\sigma^{\mathrm{AP}}$, a GS
obtained for antiperiodic BCs. One can realize
antiperiodic BCs by inverting the signs of one column of spins,
described by the Hamiltonian $H^{\mathrm{AP}}(\sigma)$.
Considering both spin configurations,  MEDWs separate
two regions on the spin lattice: one, where the spins have the same
orientation in both GSs and another, where the spins have different
orientations regarding the two GSs. In the above periodic-antiperiodic
setup, one domain wall will run through the inverted bonds. This
straight-line domain wall is not of interest to us. 
Instead, we are interested in
the MEDW which consists of those bonds that are fulfilled in exactly one
of the two GSs. 
Hence, the energy of the MEDW is given by
$H^{\mathrm{AP}}(\sigma^{\mathrm{AP}})-H(\sigma)$ and this is the minimum
energy among all the
system-spanning (top-down) DWs.
This MEDW feature is an integral part of our attempt to
study the problem of finding MEDWs in terms of a minimum-weight path problem. 

First, we will now outline the main steps of our algorithm and
elaborate on them afterwards. The algorithm can be decomposed into the
following 4 steps: For each given
realization of the bond disorder
 (i) find a GS spin configuration consistent with periodic BCs in 
$x-$direction, (ii) determine all possible MEDW segments on the dual of the
spin lattice, (iii) map  the problem to an auxiliary graph and find a
minimum weighted perfect matching (MWPM), (iv) interpret the MWPM as
minimum-weight path on the dual graph that represents a MEDW on the spin
lattice. 

Step (i): For the calculation of the GS $\sigma$, 
we apply an exact matching
algorithm that works for planar systems without external fields, e.g.\
a square lattice with periodic BCs in at most one
direction. For this purpose, the system has to be represented by its
frustrated plaquettes and paths connecting those pairwise. Finding a
minimum-weighted perfect matching on the graph of frustrated
plaquettes then corresponds to finding a GS spin configuration on the
spin lattice. For a comprehensive description, see Refs.\
\cite{barahona1982,bieche1980, opt-phys2001,janke2007}. 
This well established method yields the exact GS of the frustrated
spin-glass model and allows to explore large system sizes, easily
more than $10^5$ spins, within a
reasonable amount of computing time.  

\begin{figure}[t]
\centerline{
\includegraphics[width=0.45\linewidth]{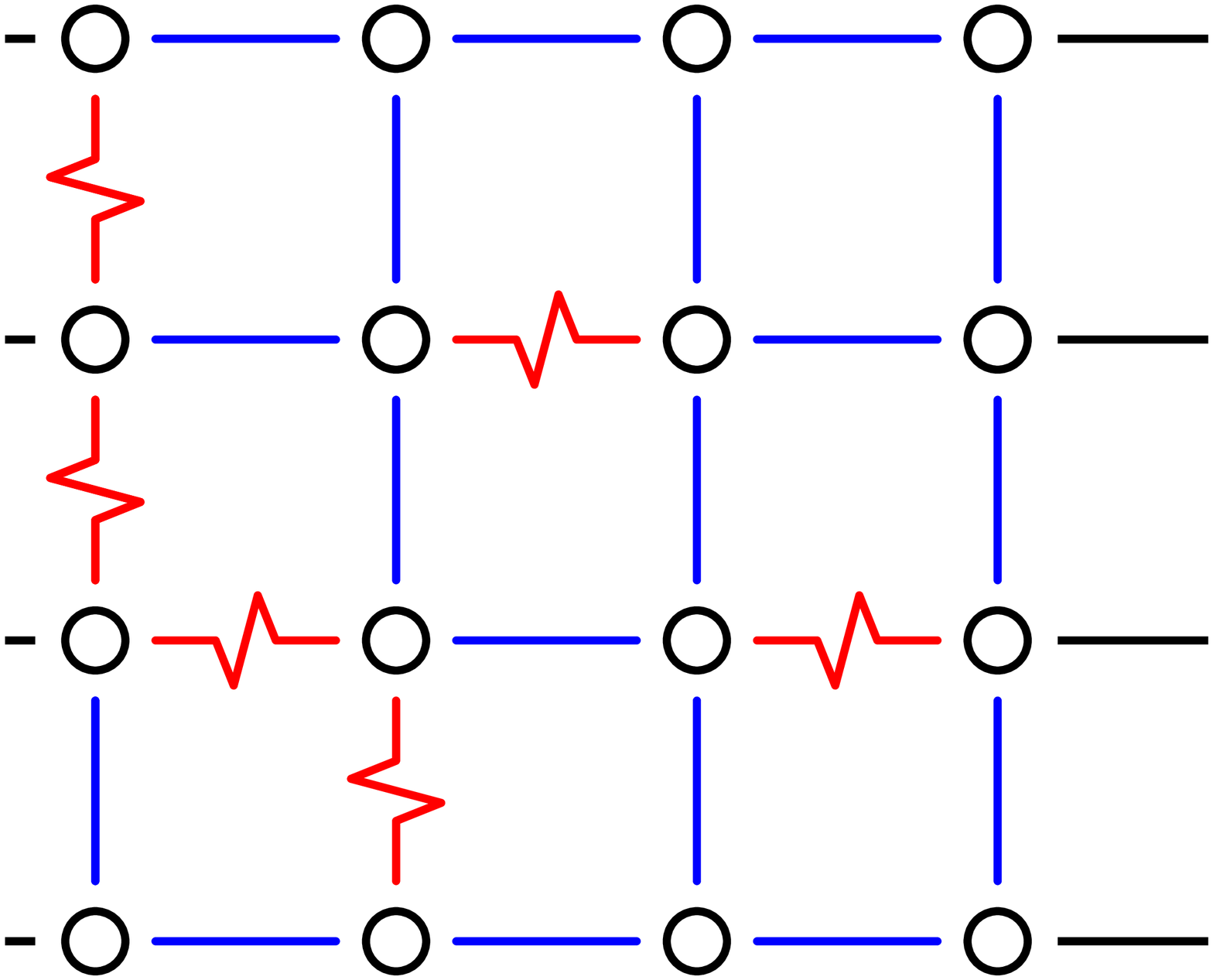}
\includegraphics[width=0.45\linewidth]{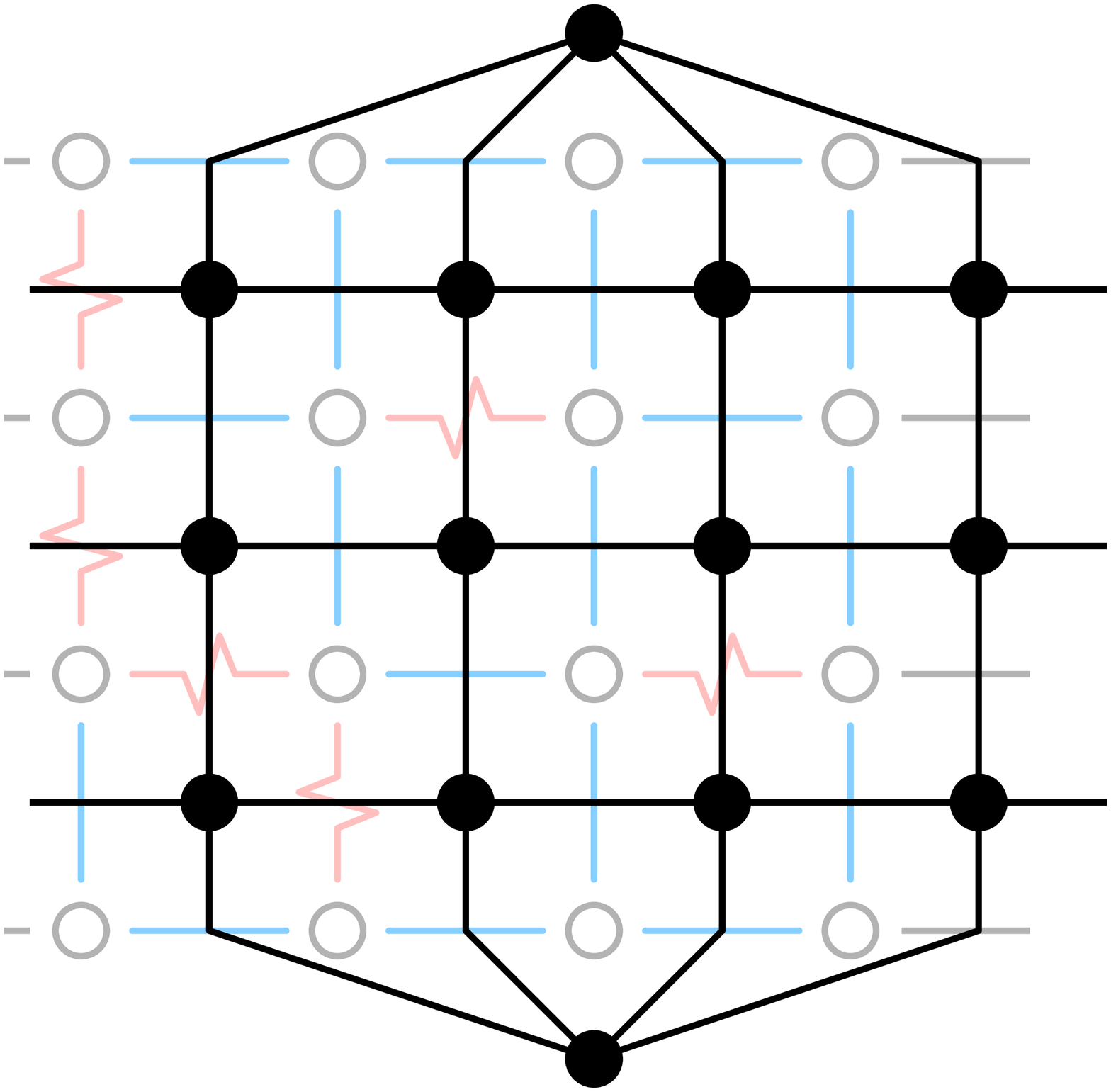}
}
\caption{(color online) Construction of the dual of the spin lattice.
Left: Spin lattice with periodic BCs in the horizontal direction
(denoted by black bonds at the left and right column).  
and free BCs along the top and bottom
boundary. The sites are denoted by the circles,
the bonds are the straight and jagged lines.
For clarity, spins are not drawn. Right: Vertices and edges
of the dual graph. Note that two extra vertices on top and at the
bottom are introduced to account for the free BCs. 
 \label{f:dual_graph}} 
\end{figure}

Step (ii): Once the GS is obtained we construct the dual of the spin
lattice as weighted graph $G = (V,E,\omega)$. The vertices $V$ are
given by all the distinct plaquettes on the spin lattice and edges
$e\in E$ connect vertices, where the corresponding plaquettes have a
bond in common, i.e.\ the edge crosses the corresponding bond, 
see Fig.\ \ref{f:dual_graph}. Note that there are two ``extra''
vertices above and below the system.
Further, a weight (or distance)
$\omega(e)$ is assigned to each edge of $G$, equal to the amount of
energy that it would contribute to a MEDW, i.e.\ 
$\omega(e)=2J_{ij}\sigma_i\sigma_j$ for $\langle i,j\rangle$ being the bond
crossed by the edge $e$, $\sigma$ being the GS obtained in step (i).
In this sense, the dual graph
comprises all possible MEDW segments. A DW is a path in the dual graph
 and the energy of a DW is the sum
of the weight of all segments being part of a DW. 
The GS property of the spin
configuration has an impact on the weight distribution of $G$. There
are negative edge weights but one cannot identify loops with negative
weight. This will be of importance
later. In summary, a  
MEDW is a path with minimum weight joining the extra
vertices of $G$. 

\begin{figure}[t]
\centerline{
\includegraphics[width=0.9\linewidth]{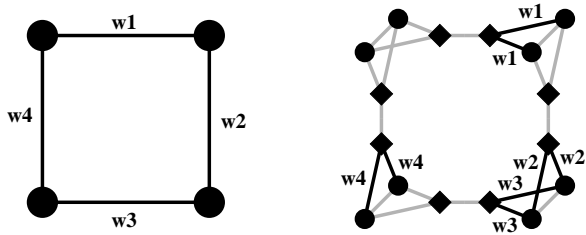}
}
\caption{
Construction of the auxiliary graph:
Every connected pair of vertices on the dual is replaced by a construct of 6 vertices and 7 edges.
This transformation is illustrated for 4 edges depicted on the left, with weights $w1$, $w2$, $w3$ and $w4$. After the transformation
the gray edges carry zero weight and the
black edges carry the same weight as the original edge on the
dual. For an illustrative purpose the vertices are divided into round
and squared ones. 
 \label{f:aux_graph}}
\end{figure}

Step (iii): Since $G$ is an undirected graph that allows for negative
edge weights, it requires matching techniques to construct minimum-weight
paths \cite{ahuja1993}. Following this reference, we 
therefore  map the problem onto an auxiliary
graph $G_A$ obtained from the dual graph by the following procedure: 
For every vertex $i$ in the dual graph (except the two extra vertices
above and below the lattice) $G_A$ contains 
a pair of ``duplicate'' vertices $i^{(a)}, i^{(b)}$
which are connected by an edge of zero weight.
Furthermore for every edge $e=(i,j)$ in the dual graph, two additional
vertices $a^{(i,j)}, b^{(i,j)}$ are introduced which are connected by
an edge $(a^{(i,j)}, b^{(i,j)})$. 
One of these two vertices is connected to the duplicate vertices
representing $i$ (e.g.\ via edges $(a^{(i,j)},i^{(a)})$,
$(a^{(i,j)},i^{(b)})$), the other one to the duplicate vertices representing
$j$ (via edges $(b^{(i,j)},j^{(a)})$,
$(b^{(i,j)},j^{(b)})$). All edge weights are zero, except for the edges
connecting one of the additional vertices to the duplicates it is
connected to, which carry the weight of $\omega(e)$, see
 Fig.\ \ref{f:aux_graph}.  We then apply an algorithm
from the LEDA \cite{mehlhorn1999,mehlhorn2000} library to obtain a
MWPM on $G_A$. 

Step (iv): The MWPM consists of a certain subset of the edges of
$G_A$. In order to interpret it as minimum-weight path on the dual graph,
one has \cite{ahuja1993}
 to perform a partial inverse transformation of step (iii). This
 means, we
merge all adjacent vertices with the same type, i.e.\ all pairs 
$i^{(a)}, i^{(b)}$ of ``duplicate'' vertices 
and all pairs $a^{(i,j)}, b^{(i,j)}$ of additional vertices.
The edges in the matching which are between vertices of the same type
disappear, while the other edges ``remain'', which means each edge 
$(a/b^{(i,j)},i/j^{(a)/(b)})$ becomes an edge $(a/b^{(i,j)},i/j)$.
Thus, some
of the edges contained in the matching disappear, while the 
remaining ones  form (for a proof and further details 
see Ref. \onlinecite{ahuja1993})
a path of minimal weight, connecting the
extra vertices of $G$. This path in turn corresponds to the MEDW on the
spin lattice. 

\begin{figure}[t]
\centerline{
\includegraphics[width=0.45\linewidth]{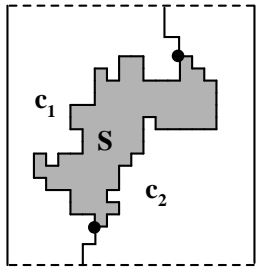}
}
\caption{{
Extension of the algorithm: subpath $c_1$, replacement path $c_2$ and
pivoting spin cluster $S$ that might be flipped in order to 
improve a MEDW regarding its length. 
} \label{f:replacement_path}}
\end{figure}  
One advantage of the above procedure, in comparison with computing the
GSs for periodic and antiperiodic boundary conditions,
is that it yields an explicit
representation of a MEDW. This allows to determine quantities like
length/roughness scaling exponents directly. The main advantage is
related to the $\pm J$--model. There, it is possible to lift the
degeneracy among the MEDWs. One can obtain
MEDWs with an exactly minimum and a maximal (i.e.\ not maximum) number of segments,
i.e.\ exactly shortest and very long MEDWs.

First, the slight modification $\omega(e) \rightarrow \omega(e) +
\epsilon$ ($\epsilon>0$)
represents a negative feedback for the inclusion of path segments,
hence the MEDW will be among all MEDWs that one which 
includes a minimal number of path segments, i.e.\ a shortest MEDW.  
The value of $\epsilon$ must be small enough to maintain the
minimum-weight path structure on the dual graph, e.g.\ 
$\epsilon < 1/|E|$.
This
modification is referred to as $\pm J$--min. The modification
$\omega(e) \rightarrow \omega(e)-\epsilon$ represents a positive
feedback for the inclusion of path segments to the MEDW. Hence, the MEDW
obtained in this way will be, among all MEDWs, one which
will include a large number of path segments. This is referred to as
$\pm J$--max. However, in the latter case arises a serious difficulty:
the weight distribution may allow for loops with negative weights now,
 and, due to the nature of the MWPM problem, the algorithm
returns a minimum-weight path in the presence of loops with negative
weight. Hence, only the total number of segments of the minimum-weight
 path together
with all loops (of zero energy in the unmodified graph)
is maximized, not the number of segments of a minimum-weight path alone.
So we are only able to obtain a lower bound on the longest
MEDWs
via using the $\pm J$-max approach. Note that obtaining the true
longest minimum-weight MEDWs is an NP-hard problem, which means that
only algorithms are known, where the running time increases
exponentially with the number $N$ of spins.

Another drawback is the fact, that the MEDW depends on the spin
configuration determined in step (i). To get really the 
shortest MEDWs ($\pm J$-min2), 
we therefore allow the algorithm
to change the GS, if this leads to a shortening of the MEDW.
The basic idea is that different GSs differ by a finite number of
zero-energy clusters of reversed spins. We are only interested in 
clusters, where part of the boundary
coincides with a MEDW, which lead to a different MEDW of the same energy 
when flipping the cluster. Note that also the edges connecting to the
two extra nodes are considered here as part of the MEDW, hence a flip of
one or several zero-energy clusters might lead to another MEDW
which has nothing in common with the starting MEDW.
Technically, we look for the shortest MEDWs 
by finding replacement paths for
certain subpaths of the MEDW, depicted in Fig.\
\ref{f:replacement_path}. Firstly, one has to find a subpath $c_1$ of
the MEDW with weight $\omega(c_1)<0$. If there is an replacement path
$c_2$ with $\omega(c_2)=|\omega(c_1)|$ and  
if $\mathrm{len}(c_2) < \mathrm{len}(c_1)$ flip the cluster of spins
surrounded by $c_1$ and $c_2$ to yield a shorter path with minimum 
weight. This is repeated until no such shortening of the MEDW is
possible any more.  
Since the flipping of the spin cluster does not cost energy (the loops
$c_1,c_2$  are zero-energy loop in the unmodified graph), the GS
property of the spin configuration is maintained.  
This latter elaboration of the algorithm is computationally more
expensive because we have to consider all possible subpaths $c_1$ with
$\omega(c_1)<0$. This means, we can obtain the true shortest
MEDWs, considering all possible GSs, only for small sizes
$L\le 32$. For large sizes, we study only the shortest MEDW
for the given GS, which was obtained in step (i).

\begin{table}[b!]
\begin{center}
\begin{tabular}[c]{l@{\qquad}l@{\qquad}l@{\qquad}l@{\qquad}l}
\hline
\hline
             & $L\!<\!160$    & $L\!=\!160$ & $L\!=\!226$ & $L\!=\!320$    \\
\hline
Gaussian     & 40000 & 10000 & 2800  & 2000  \\ 
$\pm$J--min  & 40000 & 40000 & 20000 & 10000 \\ 
$\pm$J--max  & 40000 & 20000 & 5000  & 5900  \\ 
\hline
\hline
\end{tabular}
\end{center}
\caption{{Number of samples investigated for the different system
    sizes.}\label{t:no_samples}} 
\end{table}

\section{Results}
\label{sec:results}


\begin{table}[b!]
\begin{center}
\begin{tabular}[c]{l@{\qquad}l@{\qquad}l@{\qquad}l@{\qquad}l}
\hline
\hline
             & $d_f$    & $Q$ & $d_r$ & $Q$ \\
\hline
Gaussian     & 1.274(2) & 0.45 & 1.008(11) & 0.40 \\ 
$\pm$J--min  & 1.095(2) & 0.27 & 1.006(6) & 0.78 \\ 
$\pm$J--max  & 1.395(3) & 0.16 & 0.993(8) & 0.35 \\ 
\hline
\hline
\end{tabular}
\end{center}
\caption{Scaling exponents of the mean MEDW length ($d_f$) and its
    roughness ($d_r$).  
Fits are restricted to $L \geq 26$ ($d_f$) and $L\geq 50$ ($d_r$). The value
of $Q$ gives a measure for the quality of the fit
\cite{q_value}.\label{t:scaling_exponents}} 
\end{table}
 
\begin{figure}[t!]
\centerline{
\includegraphics[width=0.95\linewidth]{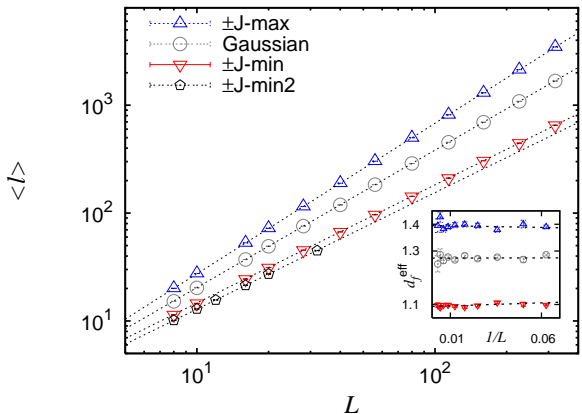}
}
\caption{(color online)
Scale dependence of the average MEDW length. Fits to the form $\langle \ell
\rangle \sim L^{d_f}$ (dashed lines) yield the exponents quoted in 
Tab.\ \ref{t:scaling_exponents}.
The inset shows the local slopes for the Gaussian and $\pm J$-min(max) MEDWs 
as function of $1/L$, providing comparable values for $d_f$ 
but with larger error bars (see text).
 \label{f:scaling_DW_length}}
\end{figure}  

%
In Fig.\ \ref{fig:sampleDWs} we show several MEDWs for a 
system with $\pm J$ disorder, one long $\pm J$-max MEDW
and one shortest $\pm J$-min MEDW, respectively. 
Already from these samples one can expect
 an observable difference in their scaling behavior. 

We expect the disorder-averaged MEDW length $\ell$ to scale with the 
system size according to $\langle \ell \rangle \sim L^{d_f}$, 
where $\langle \ldots \rangle$ denotes the disorder average and
$1\leq d_f \leq 2$ denotes the fractal dimension
of the MEDW. Its roughness $r$, 
given by the difference in the coordinates of its
leftmost and rightmost position on the spin lattice, should display the
scaling behavior $\langle r \rangle \sim L^{d_r}$, with a roughness exponent
$d_r$. 
  To determine the scaling behavior of the MEDWs, 
we have studied systems with sizes up to $L\!=\!320$ and averaged over up to
$40000$ realizations of the disorder, see Tab.\ \ref{t:no_samples}.

In Fig.\ \ref{f:scaling_DW_length}, we show the result for
$\langle\ell\rangle$ as function of system size for the Gaussian
disorder and the three cases studied for the $\pm J$ model. 
The data can be fitted very well to power laws, the results are shown in
Tab.\ \ref{t:scaling_exponents}. 
As an alternative method,
we also estimate the value of the scaling exponent $d_f$ using the local
(successive) slopes of the data points, see also \cite{buldyrev2004}.
The results are compatible with the data given in Tab.\ 
\ref{t:scaling_exponents}, only this procedure leads to a 
more conservative estimate of the numerical error, 
leading to the error bars as given in Tab.\ 
\ref{t:scaling_exponents}.

The estimate for the MEDW fractal dimension in case of a Gaussian distribution
of the interaction strengths is in good agreement with earlier results, but
has an enhanced precision, see Tab.\ \ref{t:fract_gauss}. 
Note that the treatment of large systems reduces the influence 
of systematic errors due to unknown corrections to scaling, hence providing 
a very reliable result for $d_f$.
%
Further, it is consistent with the result 
$d_f^{\mathrm{SLE}}\!=\!1.2764(4)$ predicted by the SLE
scaling relation,
where we considered $\theta\!=\!-0.287(4)$ from Ref.\ \cite{aspect-ratio2002}. 
For the $\pm J$--model we obtained a lower bound $d_f\!=\!1.095(2)$ that is
distinct from $1$, indicating that overhangs are still significant for MEDWs
with minimal length.  
Further, the estimate of the upper bound $d_f\!=\!1.395(3)$ points out that
DWs with maximal length are not space-filling. 
Note that using the past results \cite{roma2006,weigel2007}, which
are based on an uncontrolled sampling of domain walls, one could {\em not}
exclude these values $d_f=1$ and $d_f=2$.

As pointed out in the description of the algorithm, the $\pm J$--min value
actually overestimates the scaling behavior of MEDWs with minimal
length. Therefore we performed calculations with the computationally more
expensive algorithm, that allows for a change of the GS spin configuration. We
considered system sizes $L\leq 32$ with up to $3000$ samples and subsequent
fits were restricted to $12\!\leq\!L\!\leq\!32$. Albeit affected by finite
size effects, we found $d_f\!=\!1.080(5)~(Q\!=\!0.40)$ for the $\pm J$--min2
DWs, compared to a value of $d_f\!=\!1.105(2)~(Q\!=\!0.43)$ when
fitting the results for $\pm
J$--min MEDWs in the same interval.  
Hence,
minimal-length and true minimum-length MEDWs are very similar, which means
that the $\pm J$-min MEDWs, where we can obtain results for
large sizes, yield a reliable estimate of the
 behavior of shortest MEDWs.

Also, 
we analyzed the scaling behavior of MEDWs of different energies for
the $\pm J$--model, in particular for those which have energies 
$E_{\mathrm{DW}}=0$ or  $E_{\mathrm{DW}}=2$ (which constitute 96\% of
all MEDWs, the remaining 4\% are  $E_{\mathrm{DW}}=4,6,8,10$).
In all cases we find again power-law behavior (not shown) with 
$\langle \ell \rangle \sim L^{d_f}$, the resulting exponents are almost 
compatible within error bars to the average result above.
Note that in recent studies \cite{fisch2006,fisch2007} of the
DW entropy, contrarily, the behavior of $E_{\mathrm{DW}}=0$ and
$E_{\mathrm{DW}}\neq 0$ DWs was very different.


\begin{figure}[htb]
\centerline{
\includegraphics[width=0.9\linewidth]{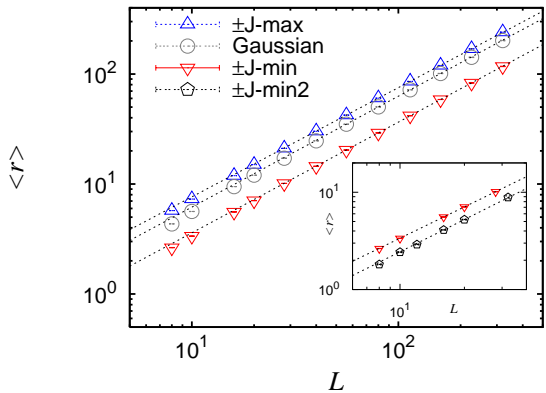}
}
\caption{{(color online)
Scale dependence of the average MEDW roughness. 
Fits to the form $\langle r
\rangle \sim L^{d_r}$ for $L \geq 50$ (dashed lines) yield exponents 
 listed in Tab.\ \ref{t:scaling_exponents}. 
The inset shows the scaling of $\pm J$--min
compared to the $\pm J$--min2 MEDWs. 
} \label{f:scaling_DW_roughness}}
\end{figure}  

Also for the scaling of the domain-wall roughness with system size, we
find a power law behavior, see Fig.\ \ref{f:scaling_DW_roughness}.
The resulting roughness exponents are shown again in Tab.\
\ref{t:scaling_exponents}.
The roughness exponents, obtained using local slopes
exhibit again larger error bars.
Hence, in all cases studied here 
the roughness exponent appears to be  
compatible with unity, showing that
the width of the MEDWs scales like the extension in $y$-direction, as
predicted in Ref. \cite{kawashima1999}. 
Regarding the roughness, one finds stronger finite-size effects for small
$L$, where we again have compared minimal-length and true minimum-length
MEDWs. Fits to the data in the interval $16\!\leq\!L\!\leq\!32$ yields
$d_r\!=\!1.101(15)~(Q\!=\!0.75)$ for the true minimum-length $\pm J$--min2 MEDWs and
$d_r\!=\!1.070(2)~(Q\!=\!0.79)$ for the minimal-length $\pm J$--min MEDWs. Hence,
minimal-length and true minimum-length MEDWs are again very similar, which means
that the $\pm J$-min MEDWs yield a reliable estimate of the
shortest MEDW behavior.

\begin{figure}[htb]
\centerline{
\includegraphics[width=0.9\linewidth]{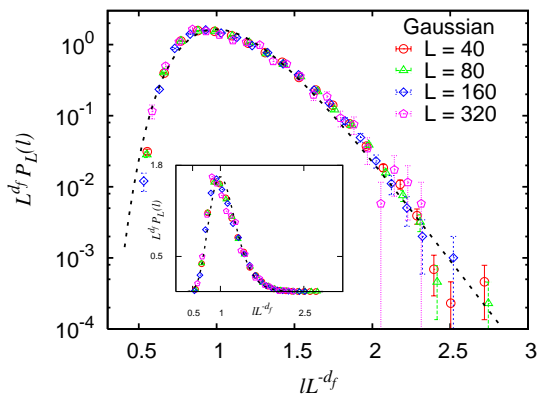}
}
\caption{{(color online)
Scaling plot of the MEDW length distribution $P_L(\ell)$ for Gaussian disorder,
where the dashed line is a log-normal distribution with mean $\mu=1.06(1)$  
and standard deviation $\sigma = 0.24(1)$, obtained from a fit to the data
of $L=160$.						
}\label{f:DW_length_distribution}} 
\end{figure}  

So far, we have analyzed the scaling behavior of mean values, now we
turn to the full distributions.
The main result is that 
the distributions $P_{L}(\ell)$ of the MEDW length for different system sizes
can be related to each other via a simple scaling relation. 
As can be seen from Fig.\ \ref{f:DW_length_distribution}, rescaling of
the length distributions according to \mbox{$P_L(\ell) = L^{-d_f}
  f(\ell L^{-d_f})$} yields a collapse to a
master curve. A qualitative similar behavior was found for the scaling
of optimal paths in strong and weak disorder
\cite{buldyrev2004,buldyrev2006}, the mass distribution of the
backbone in critical percolation \cite{barthelemy1999} and regarding
undirected minimum-weight paths in 2d lattice graphs, where the effect
of isotropically correlated bond weights on the scaling exponents was
investigated \cite{schorr2003}. 
Note that the distribution is peaked close to $\langle \ell \rangle$. 
This holds also for MEDWs subject to $\pm J$ disorder (not shown), 
where one can observe an additional even/odd deviation in the 
distribution of the MEDW lengths. 
This deviation results in a preferential appearance of domain walls with even
length, responsible for defect-energies $E_{\mathrm{DW}}=0~\rm{mod}(4)$.

As it turns out, the particular shape
of the scaling function is very simple for the case of Gaussian disorder.
It is possible to fit the distributions in this case
satisfactory by use of log-normal scaling functions. 

Fig.\ \ref{f:DW_length_distribution} shows an example, where the
log-normal distribution
\mbox{$p(x)\!=\!\exp{[-(\ln(x/\mu))^2/2\sigma^2]}/(x\sigma\sqrt(2\pi))$} with
$x\!=\!\ell L^{-d_f}$ was fit to the data ($L\!=\!160$), 
resulting in a mean and standard
deviation $\mu\!=\!1.06(1)$ and $\sigma\!=\!0.24(1)$, respectively. 
This scaling function does not suit the bimodal disorder case. Here, for MEDWs with minimal
length, we observe an exponential decay of $P_L(\ell)$ with
increasing $\ell \geq \langle \ell \rangle$ and estimated the decay exponent $\alpha=-5.9(2)$ from a fit to the data corresponding to $L=320$.  
The DWs with maximal length somehow resemble the Gaussian case, but we
could not find a meaningful distribution to describe it. 


\begin{figure}[htb]
\centerline{
\includegraphics[width=0.9\linewidth]{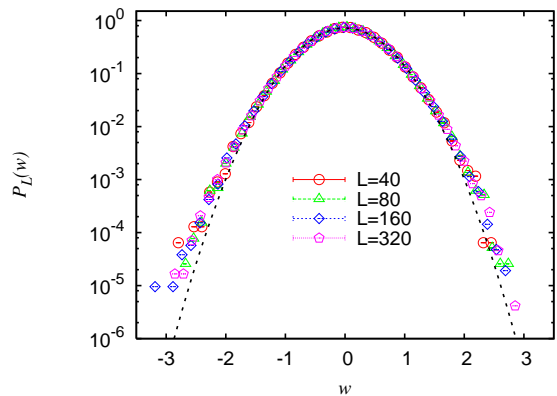}
}
\caption{{(color online)
Distribution $P_L(w)$ of weights $w$ on the MEDW segments for different system
sizes $L$. 
The dashed line is a Gaussian distribution with 
mean $\mu=-0.0003(40)$ and standard deviation $\sigma=0.552(3)$, 
obtained from a fit to the data corresponding to $L=320$. 
Results where averaged over $10^{3}$ samples.
} \label{f:wgt_distr_sp_gauss}}
\end{figure}  

Finally, we look at
the distribution $P_L(w)$ of weights $w$ which comprise the segments along the
obtained MEDW for
different system sizes, see Fig.\ \ref{f:wgt_distr_sp_gauss}.
Since the segments available to the DWs are a result of a GS
calculation, and because 
each MEDW is the
result of another optimization procedure, the behavior is a priori not clear. 
Clearly, bonds with very negative energy will not occur, because these
bonds with a large absolute value will be satisfied in the ground state
with high probability, hence yield a positive contribution to the
energy of a domain wall. 
This contributes to a decrease of the width of the distribution $P_L(w)$, 
compared to the underlying disorder distribution. 
Apart from this effect, the resulting data compares well to a Gaussian 
with mean 0 and a standard deviation close to 0.5: from a 
 fit to the data at $L=320$, we have obtained a mean
$\mu=-0.0003(40)$  and a standard deviation $\sigma=0.552(3)$. Note that the
distributions do not seem to depend strongly on the system size.


\section{Summary}
In summary, we have performed GS and DW calculations for 2d Ising spin
glasses with short ranged interactions via exact optimization
algorithm. Exploring large system sizes, we investigated the fractal
properties of MEDWs arising from Gaussian and bimodal disorder.

Our approach is based on a minimum-weight path approach for paths
on undirected networks which can have negative edge weights. This
allows for a more direct calculation of MEDWs.
Presently, in the case of the degenerate $\pm J$ model,
 we are able to calculate the shortest and very long paths,
which allows to obtain bounds on the fractal properties of
the MEDWs. We believe that this approach could severe as a basis
for the desired calculation of {\em typical} DWs for the $\pm J$ model.

 For the $\pm J$ model we found a lower bound 
$d_f\!=\!1.095(2)$, which is clearly larger than unity, 
and we estimated 
an upper bound $d_f\!=\!1.395(3)<2$. These exponents do not change when
we analyze the scaling behavior of MEDWs restricted to certain
energies. 
 Our
results for the length scaling exponent in case of Gaussian disorder
$d_f\!=\!1.274(2)$ is in agreement with prior results but with
strongly enhanced
precision. Further, it compares well to the result obtained from the
recently proposed SLE scaling relation.
Furthermore, we found that the full distributions of the MEDW lengths
scale with the same fractal exponent $d_f$. This behavior was also found
for observables in a different physical context. Finally, the width of
the domain wall scales like its height, for all cases considered here.

For later work, one could consider models with a $T=0$ transition
between a ferromagnetic phase and a spin-glass phase, and study the
behavior of the fractal dimensions around this transition. Here, from
Eq. (\ref{eq:exponents}), one
would expect to observe a discontinuous change of the exponents when
approaching the transition, because $\theta$ is expected to be the same
everywhere in the spin-glass phase.
Furthermore, it would be desirable to be able to sample GSs for the
$\pm J$ model in equilibrium, hence each GS would contribute provably with
equal weight/probability to all results. This would allow to determine
a fractal dimension of the average domain wall. This value might be
related to the energy-scaling exponent $\theta$ via a relation similar
to Eq.\ (\ref{eq:exponents}). 

\begin{acknowledgments}
We would like to thank Ron Fisch, Matthew Hastings and Mike Moore 
for many valuable discussions. 
Furthermore, we are grateful to Mike Moore for a critical reading of the 
manuscript. 
 We obtained financial support from the VolkswagenStiftung (Germany)
 within the program ``Nachwuchsgruppen an Universit\"aten''. Financial support
by the European Community's Human Potential
 Program under contract number HPRN-CT-2002-00307, DYGLAGEMEM 
 is also acknowledged. 
 The simulations were performed at  the ``Gesellschaft f\"ur Wissenschaftliche
Datenverarbeitung'' and the workstation cluster of the ``Institute for
Theoretical Physics'', both in G\"ottingen, Germany.  
\end{acknowledgments}         


\bibliography{fract_dim_dw.bib}

\end{document}